\RequirePackage[left]{lineno}
\documentclass[aps,prd,superscriptaddress,floatfix,nofootinbib,reprint]{revtex4-1}

\usepackage{graphicx}
\usepackage{amsmath}
\usepackage{longtable}
\usepackage{aas_macros}
\usepackage{xcolor}
\usepackage{multirow}
\usepackage{mathrsfs}
\usepackage[utf8]{inputenc} 

\graphicspath{{./figures/}}

\usepackage[sort&compress]{natbib}
\usepackage{hyperref}
\usepackage{url}
\newcommand{\newText}{\textcolor{black}}  

\begin{document}

\title{Search for Gamma-ray Spectral Lines from Dark Matter Annihilation in Dwarf Galaxies with the High-Altitude Water Cherenkov Observatory}

\date{\today}


\author{A.~Albert}
\affiliation{Physics Division, Los Alamos National Laboratory, Los Alamos, NM, USA }
\email{amalbert@lanl.gov}
\author{R.~Alfaro}
\affiliation{Instituto de F\'{i}sica, Universidad Nacional Autónoma de México, Ciudad de Mexico, Mexico }
\author{C.~Alvarez}
\affiliation{Universidad Autónoma de Chiapas, Tuxtla Gutiérrez, Chiapas, México}
\author{J.C.~Arteaga-Velázquez}
\affiliation{Universidad Michoacana de San Nicolás de Hidalgo, Morelia, Mexico }
\author{K.P.~Arunbabu}
\affiliation{Instituto de Geof\'{i}sica, Universidad Nacional Autónoma de México, Ciudad de Mexico, Mexico }
\author{D.~Avila Rojas}
\affiliation{Instituto de F\'{i}sica, Universidad Nacional Autónoma de México, Ciudad de Mexico, Mexico }
\author{H.A.~Ayala Solares}
\affiliation{Department of Physics, Pennsylvania State University, University Park, PA, USA }
\author{E.~Belmont-Moreno}
\affiliation{Instituto de F\'{i}sica, Universidad Nacional Autónoma de México, Ciudad de Mexico, Mexico }
\author{S.Y.~BenZvi}
\affiliation{Department of Physics \& Astronomy, University of Rochester, Rochester, NY , USA }
\author{C.~Brisbois}
\affiliation{Department of Physics, University of Maryland, College Park, MD, USA }
\author{K.S.~Caballero-Mora}
\affiliation{Universidad Autónoma de Chiapas, Tuxtla Gutiérrez, Chiapas, México}
\author{T.~Capistrán}
\affiliation{Instituto Nacional de Astrof\'{i}sica, Óptica y Electrónica, Puebla, Mexico }
\author{A.~Carramiñana}
\affiliation{Instituto Nacional de Astrof\'{i}sica, Óptica y Electrónica, Puebla, Mexico }
\author{S.~Casanova}
\affiliation{Institute of Nuclear Physics Polish Academy of Sciences, PL-31342 IFJ-PAN, Krakow, Poland }
\author{U.~Cotti}
\affiliation{Universidad Michoacana de San Nicolás de Hidalgo, Morelia, Mexico }
\author{J.~Cotzomi}
\affiliation{Facultad de Ciencias F\'{i}sico Matemáticas, Benemérita Universidad Autónoma de Puebla, Puebla, Mexico }
\author{S.~Coutiño de León}
\affiliation{Instituto Nacional de Astrof\'{i}sica, Óptica y Electrónica, Puebla, Mexico }
\author{E.~De la Fuente}
\affiliation{Departamento de F\'{i}sica, Centro Universitario de Ciencias Exactase Ingenierias, Universidad de Guadalajara, Guadalajara, Mexico }
\author{S.~Dichiara}
\affiliation{Instituto de Astronom\'{i}a, Universidad Nacional Autónoma de México, Ciudad de Mexico, Mexico }
\author{B.L.~Dingus}
\affiliation{Physics Division, Los Alamos National Laboratory, Los Alamos, NM, USA }
\author{M.A.~DuVernois}
\affiliation{Department of Physics, University of Wisconsin-Madison, Madison, WI, USA }
\author{J.C.~Díaz-Vélez}
\affiliation{Departamento de F\'{i}sica, Centro Universitario de Ciencias Exactase Ingenierias, Universidad de Guadalajara, Guadalajara, Mexico }
\author{K.~Engel}
\affiliation{Department of Physics, University of Maryland, College Park, MD, USA }
\author{C.~Espinoza}
\affiliation{Instituto de F\'{i}sica, Universidad Nacional Autónoma de México, Ciudad de Mexico, Mexico }
\author{N.~Fraija}
\affiliation{Instituto de Astronom\'{i}a, Universidad Nacional Autónoma de México, Ciudad de Mexico, Mexico }
\author{A.~Galván-Gámez}
\affiliation{Instituto de Astronom\'{i}a, Universidad Nacional Autónoma de México, Ciudad de Mexico, Mexico }
\author{J.A.~García-González}
\affiliation{Instituto de F\'{i}sica, Universidad Nacional Autónoma de México, Ciudad de Mexico, Mexico }
\author{F.~Garfias}
\affiliation{Instituto de Astronom\'{i}a, Universidad Nacional Autónoma de México, Ciudad de Mexico, Mexico }
\author{M.M.~González}
\affiliation{Instituto de Astronom\'{i}a, Universidad Nacional Autónoma de México, Ciudad de Mexico, Mexico }
\author{J.A.~Goodman}
\affiliation{Department of Physics, University of Maryland, College Park, MD, USA }
\author{J.P.~Harding}
\affiliation{Physics Division, Los Alamos National Laboratory, Los Alamos, NM, USA }
\author{S.~Hernandez}
\affiliation{Instituto de F\'{i}sica, Universidad Nacional Autónoma de México, Ciudad de Mexico, Mexico }
\author{B.~Hona}
\affiliation{Department of Physics, Michigan Technological University, Houghton, MI, USA }
\author{D.~Huang}
\affiliation{Department of Physics, Michigan Technological University, Houghton, MI, USA }
\author{P.~Huentemeyer}
\affiliation{Department of Physics, Michigan Technological University, Houghton, MI, USA }
\author{F.~Hueyotl-Zahuantitla}
\affiliation{Universidad Autónoma de Chiapas, Tuxtla Gutiérrez, Chiapas, México}
\author{A.~Iriarte}
\affiliation{Instituto de Astronom\'{i}a, Universidad Nacional Autónoma de México, Ciudad de Mexico, Mexico }
\author{V.~Joshi}
\affiliation{Erlangen Centre for Astroparticle Physics, Friedrich-Alexander-Universit{\"a}t Erlangen-N{\"u}rnberg, Erlangen, Germany}
\author{A.~Lara}
\affiliation{Instituto de Geof\'{i}sica, Universidad Nacional Autónoma de México, Ciudad de Mexico, Mexico }
\author{H.~León Vargas}
\affiliation{Instituto de F\'{i}sica, Universidad Nacional Autónoma de México, Ciudad de Mexico, Mexico }
\author{J.T.~Linnemann}
\affiliation{Department of Physics and Astronomy, Michigan State University, East Lansing, MI, USA }
\author{A.L.~Longinotti}
\affiliation{Instituto Nacional de Astrof\'{i}sica, Óptica y Electrónica, Puebla, Mexico }
\author{G.~Luis-Raya}
\affiliation{Universidad Politecnica de Pachuca, Pachuca, Hgo, Mexico }
\author{J.~Lundeen}
\affiliation{Department of Physics and Astronomy, Michigan State University, East Lansing, MI, USA }
\author{K.~Malone}
\affiliation{Physics Division, Los Alamos National Laboratory, Los Alamos, NM, USA }
\author{S.S.~Marinelli}
\affiliation{Department of Physics and Astronomy, Michigan State University, East Lansing, MI, USA }
\author{O.~Martinez}
\affiliation{Facultad de Ciencias F\'{i}sico Matemáticas, Benemérita Universidad Autónoma de Puebla, Puebla, Mexico }
\author{I.~Martinez-Castellanos}
\affiliation{Department of Physics, University of Maryland, College Park, MD, USA }
\author{J.~Martínez-Castro}
\affiliation{Centro de Investigaci\'on en Computaci\'on, Instituto Polit\'ecnico Nacional, M\'exico City, M\'exico.}
\author{J.A.~Matthews}
\affiliation{Dept of Physics and Astronomy, University of New Mexico, Albuquerque, NM, USA }
\author{P.~Miranda-Romagnoli}
\affiliation{Universidad Autónoma del Estado de Hidalgo, Pachuca, Mexico }
\author{J.A.~Morales-Soto}
\affiliation{Universidad Michoacana de San Nicolás de Hidalgo, Morelia, Mexico }
\author{E.~Moreno}
\affiliation{Facultad de Ciencias F\'{i}sico Matemáticas, Benemérita Universidad Autónoma de Puebla, Puebla, Mexico }
\author{M.~Mostafá}
\affiliation{Department of Physics, Pennsylvania State University, University Park, PA, USA }
\author{A.~Nayerhoda}
\affiliation{Institute of Nuclear Physics Polish Academy of Sciences, PL-31342 IFJ-PAN, Krakow, Poland }
\author{L.~Nellen}
\affiliation{Instituto de Ciencias Nucleares, Universidad Nacional Autónoma de Mexico, Ciudad de Mexico, Mexico }
\author{M.~Newbold}
\affiliation{Department of Physics and Astronomy, University of Utah, Salt Lake City, UT, USA }
\author{M.U.~Nisa}
\affiliation{Department of Physics and Astronomy, Michigan State University, East Lansing, MI, USA }
\author{R.~Noriega-Papaqui}
\affiliation{Universidad Autónoma del Estado de Hidalgo, Pachuca, Mexico }
\author{A.~Peisker}
\affiliation{Department of Physics and Astronomy, Michigan State University, East Lansing, MI, USA }
\author{E.G.~Pérez-Pérez}
\affiliation{Universidad Politecnica de Pachuca, Pachuca, Hgo, Mexico }
\author{C.D.~Rho}
\affiliation{Department of Physics \& Astronomy, University of Rochester, Rochester, NY , USA }
\author{D.~Rosa-González}
\affiliation{Instituto Nacional de Astrof\'{i}sica, Óptica y Electrónica, Puebla, Mexico }
\author{M.~Rosenberg}
\affiliation{Department of Physics, Pennsylvania State University, University Park, PA, USA }
\author{H.~Salazar}
\affiliation{Facultad de Ciencias F\'{i}sico Matemáticas, Benemérita Universidad Autónoma de Puebla, Puebla, Mexico }
\author{F.~Salesa Greus}
\affiliation{Institute of Nuclear Physics Polish Academy of Sciences, PL-31342 IFJ-PAN, Krakow, Poland }
\author{A.~Sandoval}
\affiliation{Instituto de F\'{i}sica, Universidad Nacional Autónoma de México, Ciudad de Mexico, Mexico }
\author{M.~Schneider}
\affiliation{Department of Physics, University of Maryland, College Park, MD, USA }
\author{R.W.~Springer}
\affiliation{Department of Physics and Astronomy, University of Utah, Salt Lake City, UT, USA }
\author{E.~Tabachnick}
\affiliation{Department of Physics, University of Maryland, College Park, MD, USA }
\author{O.~Tibolla}
\affiliation{Universidad Politecnica de Pachuca, Pachuca, Hgo, Mexico }
\author{K.~Tollefson}
\affiliation{Department of Physics and Astronomy, Michigan State University, East Lansing, MI, USA }
\author{I.~Torres}
\affiliation{Instituto Nacional de Astrof\'{i}sica, Óptica y Electrónica, Puebla, Mexico }
\author{R.~Torres-Escobedo}
\affiliation{Departamento de F\'{i}sica, Centro Universitario de Ciencias Exactase Ingenierias, Universidad de Guadalajara, Guadalajara, Mexico }
\author{T.~Weisgarber}
\affiliation{Department of Physics, University of Wisconsin-Madison, Madison, WI, USA }
\author{J.~Wood}
\affiliation{Department of Physics, University of Wisconsin-Madison, Madison, WI, USA }
\author{A.~Zepeda}
\affiliation{Physics Department, Centro de Investigacion y de Estudios Avanzados del IPN, Mexico City, DF, Mexico }
\author{H.~Zhou}
\affiliation{Physics Division, Los Alamos National Laboratory, Los Alamos, NM, USA }
\author{C.~de León}
\affiliation{Universidad Michoacana de San Nicolás de Hidalgo, Morelia, Mexico }

\begin{abstract}
Local dwarf spheroidal galaxies (dSphs) are nearby dark-matter dominated systems, making them excellent targets for searching for gamma rays from particle dark matter interactions. If dark matter annihilates or decays directly into two gamma rays (or a gamma ray and a neutral particle), a monochromatic spectral line is created. At TeV energies, no other processes are expected to produce spectral lines, making this a very clean indirect dark matter search channel. With the development of event-by-event energy reconstruction, we can now search for spectral lines with the High Altitude Water Cherenkov (HAWC) Observatory. HAWC is a wide field of view survey instrument located in central Mexico that observes gamma rays from $<$1 TeV to $>$100 TeV. In this work we present results from a recent search for spectral lines from local, dark matter dominated dwarf galaxies using 1038 days of HAWC data. We also present updated limits on several continuum channels that were reported in a previous publication. Our gamma-ray spectral line limits are the most constraining obtained so far from 20 TeV to 100 TeV.

\end{abstract}
\maketitle

\section{INTRODUCTION}\label{sec:intro}

Several pieces of observational evidence suggest the majority of the matter in the Universe is composed of dark matter (DM)~\cite{Ade:2013zuv,Clowe:2006eq,Sofue:2000jx}. Many theories predict DM is composed of fundamental particles of which Weak\newText{l}y Interacting Massive Particles (WIMPs)~\cite{Feng:2010gw} are the most promising. These particles typically annihilate or decay to Standard Model particles that cascade and produce stable secondary particles like gamma rays. While most annihilation or decay channels (e.g.$\chi\chi\longrightarrow b\bar{b}$) produce a continuum of gamma rays, if the DM annihilated directly into a gamma ray and a second neutral particle (X) it would produce gamma rays with a specific energy depending on the mass of the DM particle ($m_{\rm DM}$) and the second particle ($m_{X}$):

\begin{equation}
    E_{\gamma} = m_{\rm DM} \left( 1-\frac{m^{2}_{X}}{4 m^{2}_{\rm DM}} \right).
\end{equation}

This is derived using energy conservation and assuming that the DM is cold (non-relativistic). Also we used the standard particle physics convention of $c=1$. 

These DM annihilations produce a monochromatic spectral emission line in the gamma-ray energy spectrum. This is a sharp spectral feature as opposed to the continuum emission expected from most other annihilations, for example annihilations to a pair of b quarks. For annihilation, if the second particle is another gamma ray, then the spectral line energy would be the mass of the DM particle, giving us both evidence of DM interactions and information about the particle nature of DM. For TeV gamma rays, no other process is predicted to produce a spectral line making this a clean DM search channel. However, the process is predicted to be heavily loop suppressed, with a branching fraction $\sim10^{-4}$~\cite{Bergstrom:1997fh,Ferrer:2006hy,Gustafsson:2007pc,Profumo:2008yg}. 

Indirect DM searches like the one presented here are aimed to detect gamma rays from DM interactions in cosmic sources. Therefore dwarf spheroidal galaxies (dSphs) in the Milky Way dark matter halo are one of the most promising targets for indirect DM searches given their proximity and high DM content~\cite{Strigari:2018utn}. 
Previous searches for TeV gamma rays from dSphs by the High Energy Stereoscopic System (H.E.S.S.), the Very Energetic Radiation Imaging Telescope Array system (VERITAS), the Major Atmospheric Gamma-ray Imaging Cherenkov Observatory (MAGIC), and HAWC have resulted in null detections. There were both searches for continuum emission~\cite{Ahnen:2016qkx,Aleksic:2013xea,Archambault:2017wyh,Abramowski:2014tra,Ahnen:2017pqx,Albert:2017vtb} and spectral lines~\cite{Archambault:2017wyh,Abdalla:2018mve}. H.E.S.S. has also looked for spectral lines in the inner Galactic Halo~\cite{Abdallah:2018qtu}.

\section{DATA AND ANALYSIS}\label{sec:data}

Here we search for spectral lines from 1 to 100 TeV from 11 dSphs using 1038 days of data from HAWC. We also present a search for continuum emission from dSphs in App.~\ref{sec:app} that is an update from the limits presented in Ref~\cite{Albert:2017vtb}. With its wide field of view, HAWC observes 2/3 of the gamma-ray sky each day. It detects gamma rays with energies from $<$1 TeV to $>$100 TeV. HAWC consists of 300 light-tight water Cherenkov detectors equipped with 4 photomultiplier tubes (PMTs). HAWC has a $>$90\% up time, allowing it to observe both day and night. It is located in Sierra Negra, Mexico at an altitude of 4100~m at latitude 18$^{\circ}59.7'$N and longitude 97$^{\circ}18.6'$W. More information about HAWC can be found in Ref.~\cite{HAWCCrab1}.

We describe the expected flux ($\frac{d_{\phi Ann}}{dE_{\gamma}}$) from a specific dark matter annihilation model with the following equation:
\begin{footnotesize}
\begin{equation}
\frac{d\phi_{Ann}}{dE_\gamma} = \left(\frac{\langle \sigma v\rangle}{8\pi} \frac{dN_\gamma}{dE_\gamma} \frac{1}{m_{DM}^2} \right)\left(\int_{\Delta\Omega} d\Omega	\int_{\rm l.o.s.}d\ell \rho^2_{\rm DM}(\vec{\ell}) \right). \label{eq:dmfluxA}
\end{equation}
\end{footnotesize}

\noindent The first set of parentheses in Eq.~\ref{eq:dmfluxA} combines the DM particle properties. Specifically, $\langle \sigma v\rangle$ is the channel-specific annihilation cross section and $\frac{dN_\gamma}{dE_\gamma}$ is the energy distribution of gamma rays from each annihilation for that channel. For DM annihilation to 2 gamma rays, $\frac{dN_\gamma}{dE_\gamma}=2\delta(E_\gamma - m_{\rm DM})$. 

The second set of parentheses in Eq.~\ref{eq:dmfluxA}, called the ``J-factor", is derived from the integral of the DM density ($\rho_{\rm DM}$) for a given region of interest along the line of sight. The J-factor is directly proportional to the expected gamma-ray flux from DM annihilations. The size of the J-factor depends both on the DM density distribution in the object as well as its distance. Nearby DM dominated systems have the largest J-factors, and are therefore the best targets. Here we assume each dSph as a point source.

We search for spectral lines from 10 dSphs: Bootes I, Canes Venatici I, Canes Venatici II, Coma Berenices, Hercules, Leo I, Leo II, Leo IV, Segue I and Sextans. These were all included in the previous HAWC dSph analysis~\cite{Albert:2017vtb}. We use the J-factors calculated in Ref.~\cite{Albert:2017vtb}. Four dSphs used in that analysis (Ursa Major I, Ursa Major II, Ursa Minor, and Draco) were not included since they are located at unfavorable declinations resulting in zenith angles higher than $30^{\circ}$ relative to HAWC. This is because the high declination of these sources prevents the fits from converging. We note that given their high declinations where HAWC is not as sensitive, similar to the previous dSphs analysis, these dSphs would not significantly contribute to the combined limits~\cite{Albert:2017vtb}. We also exclude Triangulum II in our DM analysis due to its uncertain J-factor and the possibility of being tidally disrupted~\cite{triIIdense,triIIJ}. We do present spectral line flux limits from Triangulum II, which do not assume a specific J-factor.

To search for gamma-ray lines we perform a binned maximum likelihood analysis, with the same analysis bins described in Ref.~\cite{HAWCCrab2}. We use a 2D binning of the data. The first dimension is 9 ``fhit" bins similar to the previous HAWC analysis. The variable ``fhit" is the fraction of available PMTs hit by the airshower. The second dimension is 12 quarter decade log$_{10}(\hat{E})$ bins, resulting in 108 bins. See Tab. 1 of Ref~\cite{HAWCCrab2} for bin details. Similar to that analysis of the Crab Nebula~\cite{HAWCCrab2}, we only select events whose core is located on the main array. This means that not all analysis bins are used since several are sparsely populated. For example, the highest energy events (E$\sim$100 TeV) whose core lands on the main array overwhelmingly hit all the photomultiplier tubes (PMTs) (fhit=1), not just a small fraction like fhit = 0.1. We use the same 37 bins chosen for the Crab Nebula spectral analysis~\cite{HAWCCrab2}.

In previous HAWC analyses we binned our data in fhit bins only. The energy dispersion in each bin was very poor~\cite{HAWCCrab1} and there was no event-by-event energy reconstruction. The current HAWC analysis has 2 energy estimators for each event. One is based on a neural network ($E_{\rm{NN}}$) that uses air shower variables like core position and shower angle. The other so-called `ground parameter' is determined from the charge profile across the PMTs from the air shower ( $E_{\rm{GP}}$). For more information on the HAWC energy estimators see Ref.~\cite{HAWCCrab2}.

With the new algorithms, the HAWC energy resolution is greatly improved. We show results using $E_{\rm{NN}}$ since that estimator has a better energy resolution~\cite{HAWCCrab2}. The energy resolution is 44\% at 1 TeV and 23\% at 100 TeV. Similar results were also attained using $E_{\rm{GP}}$. To approximate the delta function in Eq.~\ref{eq:dmfluxA}, we use a gaussian with a width of 0.1 TeV.

We fit for spectral lines from 1 -- 100 TeV, which corresponds to DM annihilations directly to two gamma rays for DM masses from 1 -- 100 TeV. We chose the fit energies to be in steps of half the energy resolution.  This is because this spacing results in missing $<10$\% of the signal if the true signal were in between two fit energies~\cite{fermiLine}. We use the energy resolution derived in Ref.~\cite{HAWCCrab2} (see Figure 6 in that reference). 
This results in 25 fit energies. Specifically we fit for a spectral line at the following energies in TeV: 1, 1.43, 1.96, 2.62, 3.44, 4.45, 5.68, 7.13, 8.83, 10.80, 13.07, 15.69, 18.66, 22.02, 25.80, 30.02, 34.72, 40.01, 45.90, 52.54, 60.12, 78.20, 89.14, 101.45 TeV.

Following Ref~\cite{Albert:2017vtb}, we calculate the likelihood for a specific DM model using
\begin{equation}
\mathscr{L} = \Pi_{i,j} \frac{(B_{i,j}+S_{i,j})^{N_{i,j}}\rm{exp}[-(B_{i,j}+S_{i,j})]}{N_{i,j}!},
\end{equation}
\noindent where $B_{i,j}$ is the number of background counts, $S_{i,j}$ is the number of signal counts, and $i$ and $j$ run over the 2D analysis bins.

We define the best fit DM annihilation cross section as the value that maximizes $\mathscr{L}$. We then quantify the preference of the signal model to the background model by calculating a test statistic (TS)
\begin{equation}
TS = -2 \rm{ln} \left(\frac{\mathscr{L}_{0}}{\mathscr{L}^{max}}\right),
\end{equation}
\noindent where $\mathscr{L}_{0}$ is the likelihood from the background-only fit and $\mathscr{L}^{max}$ is the likelihood from the best fit with the signal model. We perform the likelihood analysis using the Multi-Mission Maximum Likelihood framework (3ML)~\cite{Vianello:2015wwa}.

\section{RESULTS}\label{sec:results}

The largest $TS$ in our line search was~$TS=4.47$, which occurred in Segue I at $E_{\gamma}=5.7$ TeV. When we account for the fact that we fit for lines at several different energies, the global significance post-trials is $0.45\sigma$. See Section 5B of Ref.~\cite{fermiLine} for more on calculating the global significance.

\begin{figure}[ht]
    \centering
    \includegraphics[width=0.48\textwidth]{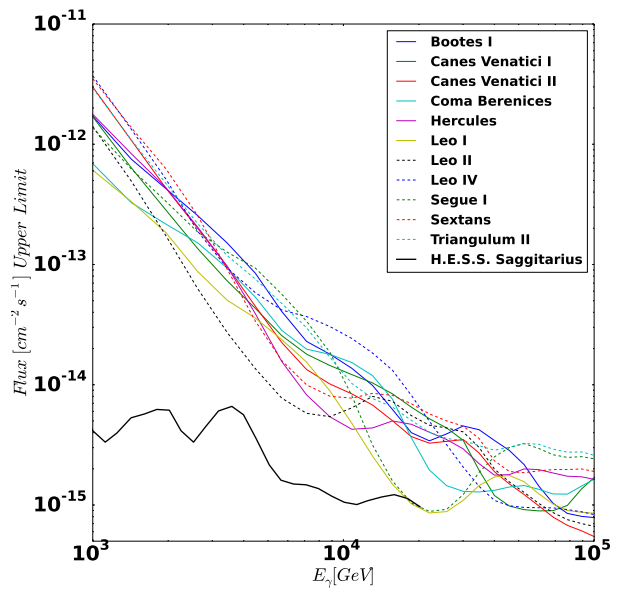}
    \caption{95\% CL upper limits on the gamma-ray spectral line flux as a function of energy for all dSphs. H.E.S.S. limits from the dSph Sagittarius are shown for comparison~\cite{Abdalla:2018mve}. }\label{fig:lineFlux}
  \end{figure}

Since we did not see any statistically significant spectral lines we will proceed to calculate 95\% confidence level (CL) upper limits~\cite{Wilks:1938dza}. Specifically, we find the energy for which the log likelihood changes by 2.71/2~\cite{pdg} relative to the best fit.
First we looked in general for spectral lines from 1 -- 100 TeV, not necessarily produced from DM interactions. The resulting limits for each dSph compared to the H.E.S.S. results from the Sagittarius dSph~\cite{Abdalla:2018mve} are shown in Fig.~\ref{fig:lineFlux}. 

\begin{figure}[ht]
    \centering
    \includegraphics[width=0.48\textwidth]{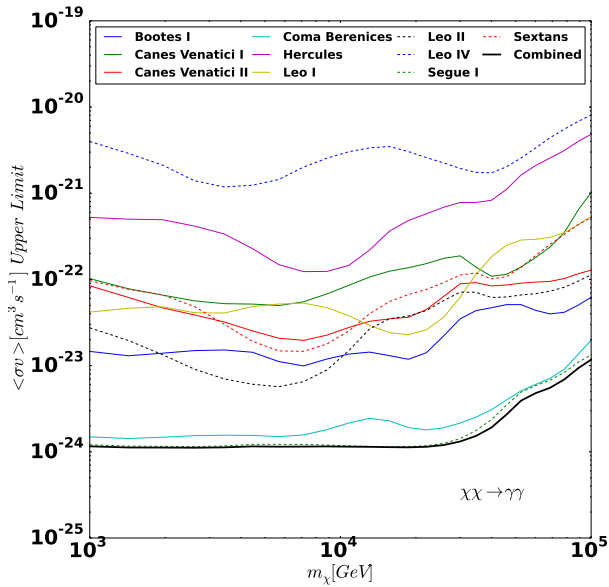}
    \caption{95\% CL upper limits on $\langle \sigma v\rangle$ for the individual dSphs. }\label{fig:lingSigV_all}
  \end{figure}

  For the DM model specific limits, we assumed DM annihilation directly to two gamma rays ($\chi\chi\rightarrow\gamma\gamma$). In this case, $m_{\rm{DM}}=E_{\gamma}$. To calculate the expected gamma-ray flux from this channel we used Eq.~\ref{eq:dmfluxA}. We fit at the 25 energy (mass) points listed in Sec.~\ref{sec:data}. Figure~\ref{fig:lingSigV_all} shows the 95\% CL upper limits on $\langle \sigma v\rangle$ for all the dSphs in this analysis.
  
\begin{figure}[ht]
    \centering
    \includegraphics[width=0.48\textwidth]{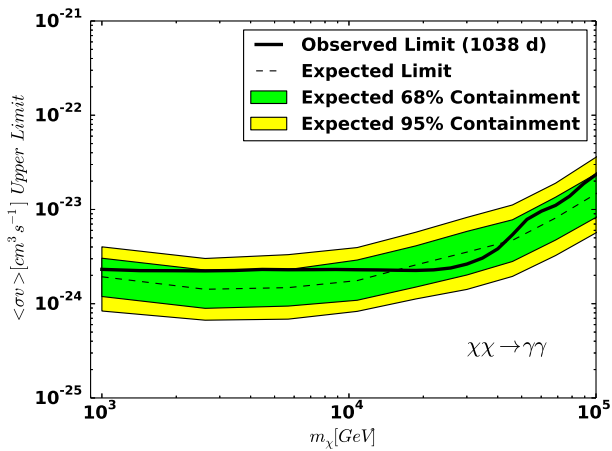}
    \caption{Combined 95\% CL upper limits on $\langle \sigma v\rangle$. Observed limits (solid black), expected limits (dashed black), and the 68\% (green band) and 95\%(yellow band) background-only containment bands are shown. Expected limits and containment bands were calculated using 1000 background-only simulations.  }\label{fig:lingSigV}
  \end{figure}

  Since we expect the same DM with the same properties in each dSph, we calculate a joint limit by combining the likelihoods of each individual dSph. Figure~\ref{fig:lingSigV} shows the observed 95\% CL upper limits on $\langle \sigma v\rangle$ along with the expectation from 1000 background only simulations. The dashed line is the median limit from those simulations and the green (yellow) band shows the 68\%(95\%) containment of the background-only limits.
  
\begin{figure}[ht]
    \centering
    \includegraphics[width=0.48\textwidth]{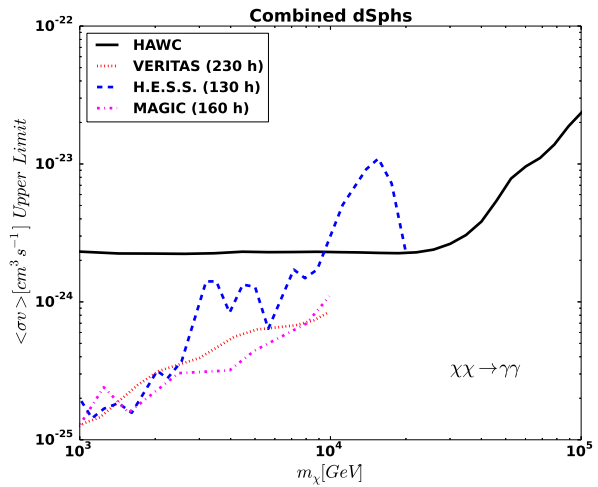}
    \caption{Joint 95\% CL upper limits on $\langle \sigma v\rangle$ compared to other experiments. VERITAS~\cite{Archambault:2017wyh}, H.E.S.S.~\cite{Abdalla:2018mve}, and MAGIC~\cite{Aleksic:2013xea} limits are from various dSph analyses (see text for details). }\label{fig:lingSigV_comp}
  \end{figure}

  Figure~\ref{fig:lingSigV_comp} compares our joint limits to those obtained by other spectral line searches in dSphs. Specifically the VERITAS result combined results from 5 dSphs (Segue I, Draco, Ursa Minor, Bootes I, and William I)~\cite{Archambault:2017wyh}, H.E.S.S. combines 5 dSphs (Fornax, Coma Berenices, Sculptor, Carina, and Sagittarius)~\cite{Abdalla:2018mve}, and MAGIC~\cite{Aleksic:2013xea}, that uses data from Segue I. Each observatory used over 100 hours of total observation time; specifically VERITAS used 230 hours, H.E.S.S. used 130 hours, and MAGIC used 160 hours.
  
\section{Discussion and Conclusions}\label{sec:conclusion}

We present a search for spectral lines from 11 dSphs from 1 to 100 TeV using 1038 days of HAWC data. The largest TS occurred at 5.7 TeV in Segue I; the TS was 4.47, corresponding to a global significance of 0.45$\sigma$. 

Since no significant spectral lines were found, we calculated 95\% confidence level upper limits to the DM annihilation cross section for the channel $\chi\chi\rightarrow\gamma\gamma$. Figure~\ref{fig:lingSigV_all} shows the DM annihilation cross section upper limits for each dSph we considered. Figure~\ref{fig:lingSigV} shows the combined limits from a joint likelihood analysis where each dSph was weighted by its J-factor. Note the combined limits are dominated by Seg I and Coma Berencies, which have the largest J-factors. We also calculated the general spectral line 95\% flux limits.

Figure~\ref{fig:lingSigV_comp} compares our combined limits with those from other searches for spectral lines from dSphs. Our limits extend the search for gamma-ray spectral lines up to 100 TeV for the first time. Specifically we provide the strongest spectral line flux limits above 20 TeV. We also have the most constraining DM spectral line annihilation cross section limits about 10 TeV from dSphs analyses.

We also show updated limits using the 1038d dataset relative to the previous 507d HAWC analysis~\cite{Albert:2017vtb} from the $b\bar{b}$, $t\bar{t}$, $\tau^{+}\tau^{-}$, $\mu^{+}\mu^{-}$, and $W^{+}W^{-}$ continuum channels in App.~\ref{sec:app}.

HAWC recently expanded its array by surrounding the main array by a large array of smaller water Cherenkov tanks called ``outriggers". These outriggers are expected to increase the HAWC effective area above 50 TeV by a factor of 3. It will allow HAWC to more accurately reconstruct the energy of high-energy showers. With the additional data from the outriggers and continued operations of HAWC, we expect to be able to extend our spectral line search up to $\sim$1 PeV in the future.

\begin{acknowledgments}

We acknowledge the support from: the US National Science Foundation (NSF); the US Department of Energy Office of High-Energy Physics; 
the Laboratory Directed Research and Development (LDRD) program of Los Alamos National Laboratory; 
Consejo Nacional de Ciencia y Tecnolog\'{\i}a (CONACyT), M{\'e}xico, grants 271051, 232656, 260378, 179588, 254964, 258865, 243290, 132197, A1-S-46288, c{\'a}tedras 873, 1563, 341, 323, Red HAWC, M{\'e}xico; 
DGAPA-UNAM grants IG100317, IN111315, IN111716-3, IA102715, IN109916, IA102917, IN112218; 
VIEP-BUAP; 
PIFI 2012, 2013, PROFOCIE 2014, 2015; 
the University of Wisconsin Alumni Research Foundation; 
the Institute of Geophysics, Planetary Physics, and Signatures at Los Alamos National Laboratory; 
Polish Science Centre, grants DEC-2018/31/B/ST9/01069, DEC-2017/27/B/ST9/02272; 
Coordinaci{\'o}n de la Investigaci{\'o}n Cient\'{\i}fica de la Universidad Michoacana; Royal Society - Newton Advanced Fellowship 180385; 
Thanks to Scott Delay, Luciano D\'{\i}az and Eduardo Murrieta for technical support.

\end{acknowledgments}
\clearpage

\appendix
\begin{widetext}
\section{Continuum Limits}\label{sec:app}

In addition to a search for spectral lines from DM annihilation, we also fit for several DM annihilation channels that create a broader continuum of gamma-rays. We show the 95\% confidence level upper limits for DM annihilation for the $b\bar{b}$, $t\bar{t}$, $\tau^{+}\tau^{-}$, $\mu^{+}\mu^{-}$, and $W^{+}W^{-}$ channels in Fig~\ref{fig:indCh}. Note the sensitivity at lower masses for this work is less than the previous 507d analysis~\cite{Albert:2017vtb} since in this analysis we make a cut that requires the air shower core center to be in the main array to better estimate the energy. The dips around $m_{\rm{DM}}=20$ TeV in the $\tau^{+}\tau^{-}$, $\mu^{+}\mu^{-}$, and $W^{+}W^{-}$ are from a $\approx 2 \sigma$  underfluctuation in Segue I that can also be seen in Figure~\ref{fig:lineFlux}.

\begin{figure*}[ht]
    \centering
    \includegraphics[width=0.45\textwidth]{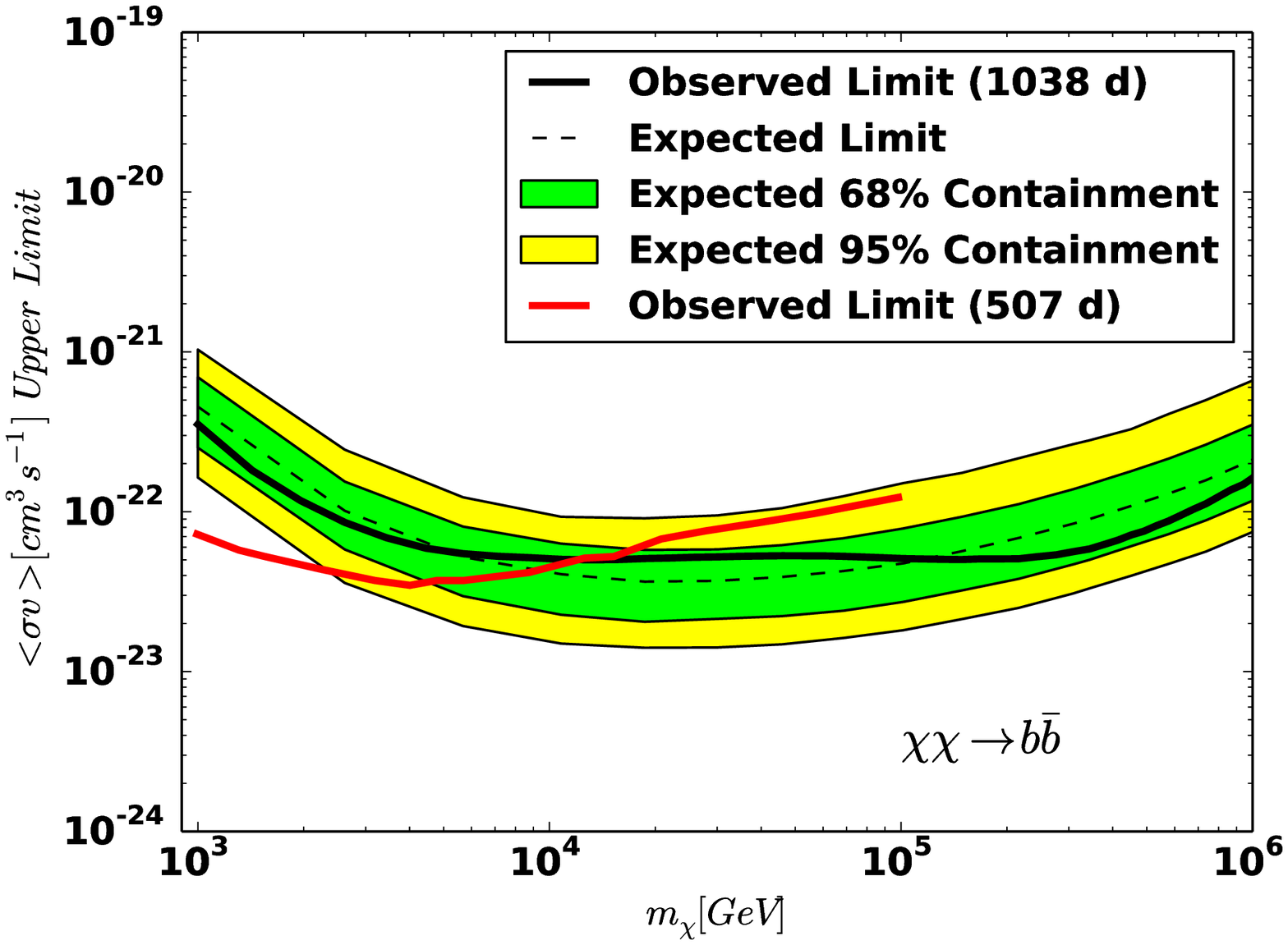}
    \includegraphics[width=0.45\textwidth]{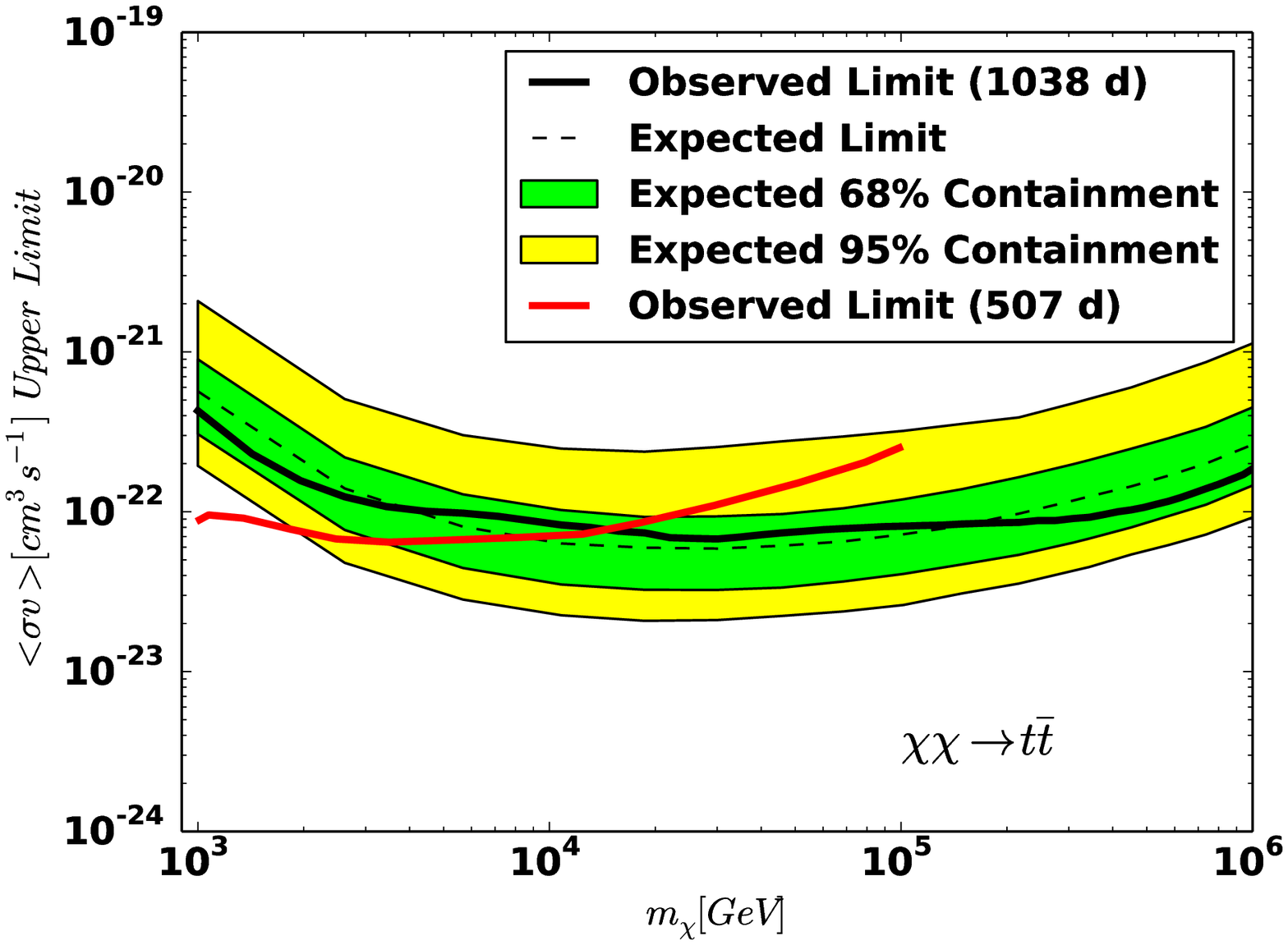}
    \includegraphics[width=0.45\textwidth]{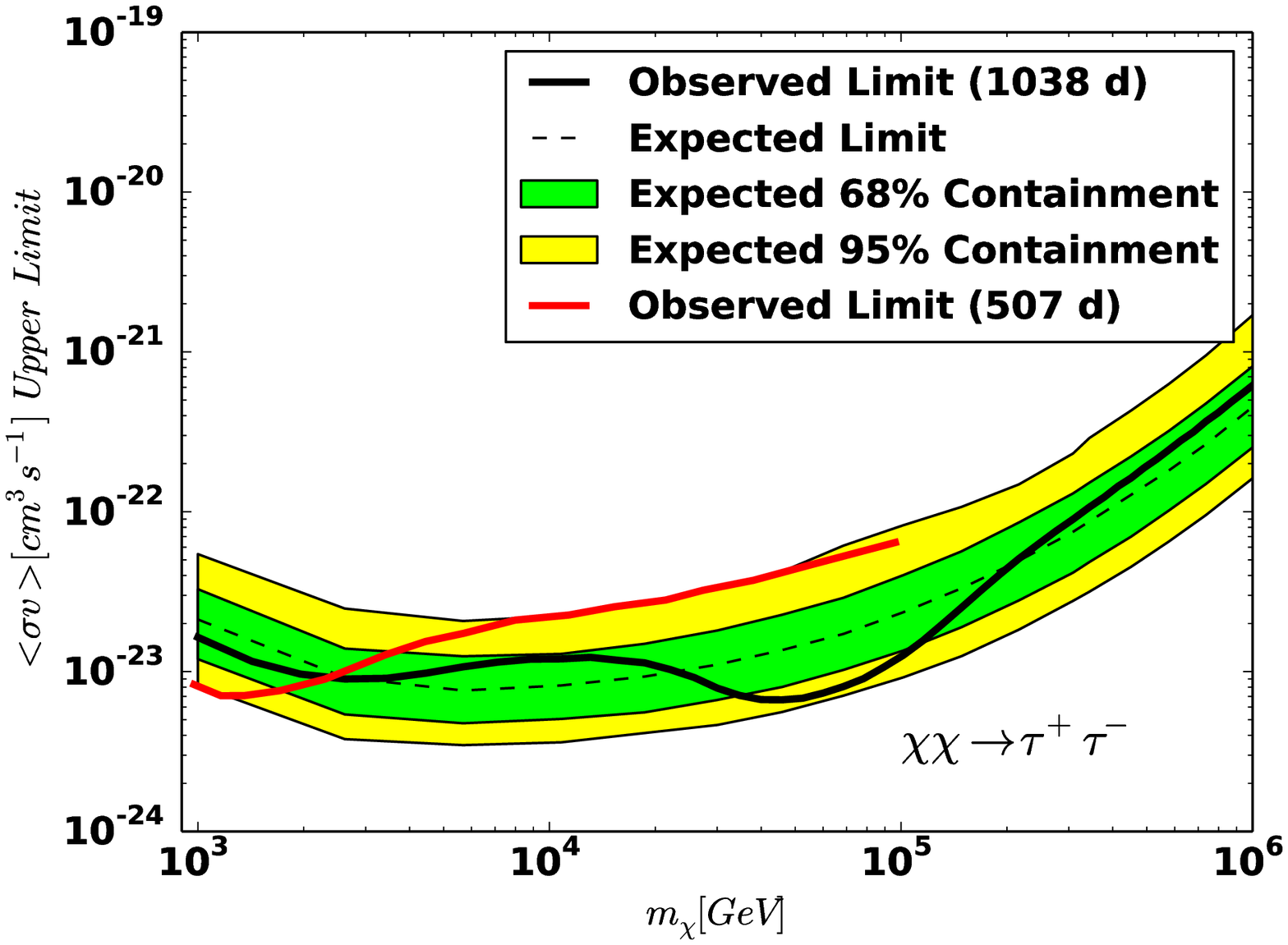}
    \includegraphics[width=0.45\textwidth]{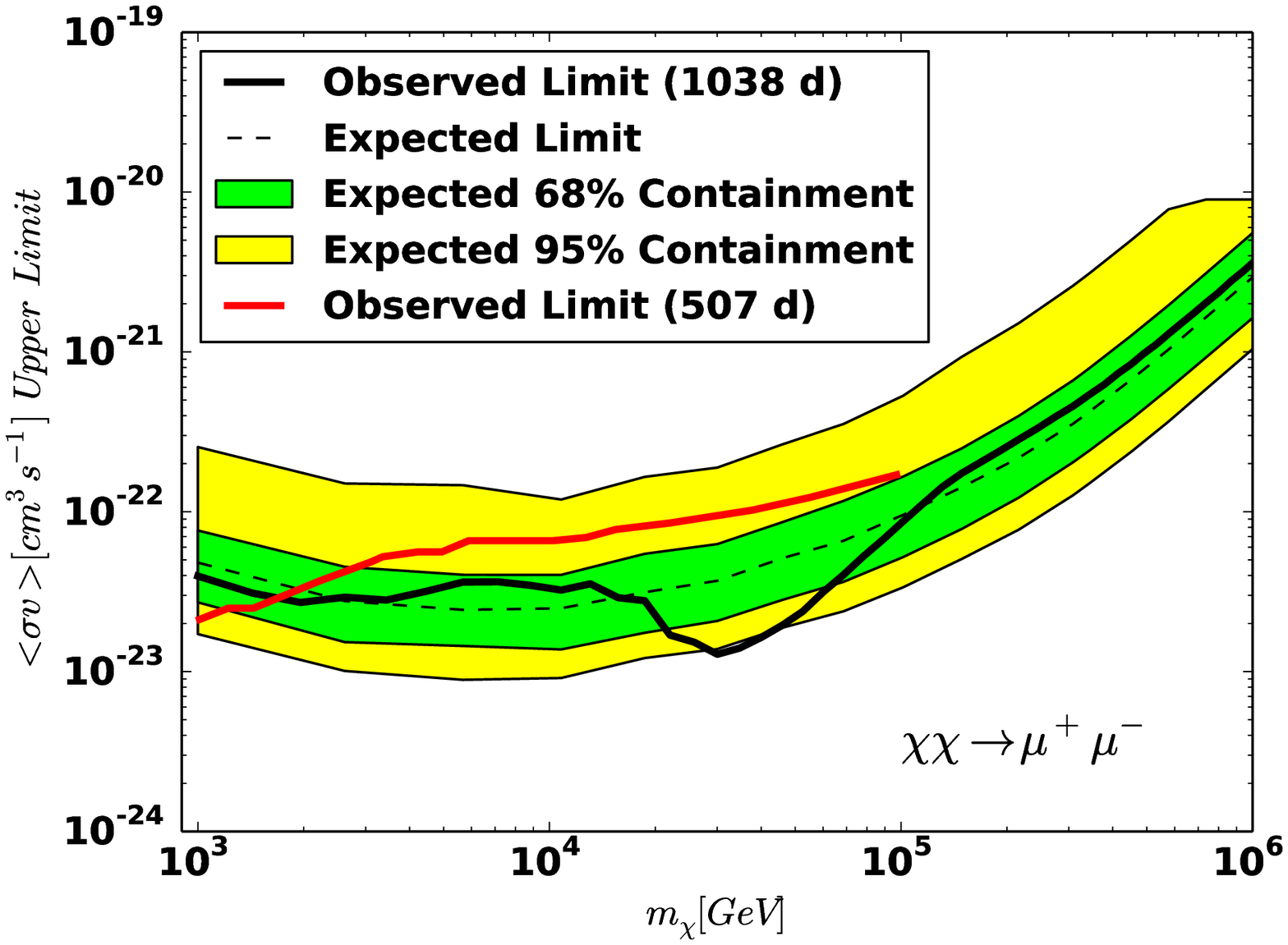}
    \includegraphics[width=0.45\textwidth]{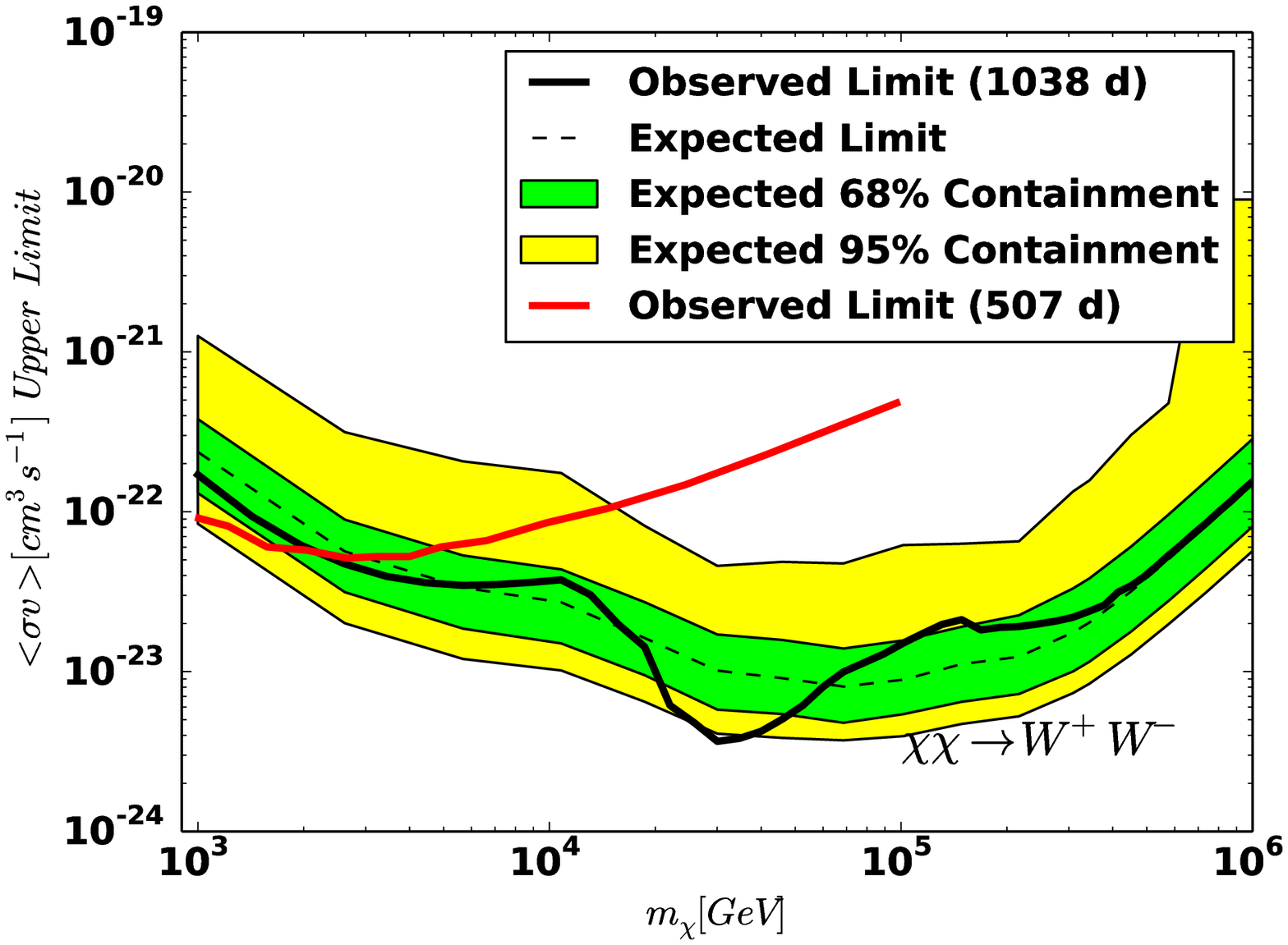}
    \caption{Combined dSphs DM $\langle \sigma v\rangle$ 95\% confidence level upper limits for the $b\bar{b}$, $t\bar{t}$, $\tau^{+}\tau^{-}$, $\mu^{+}\mu^{-}$, and $W^{+}W^{-}$ channels.  \label{fig:indCh}}
  \end{figure*}
 \pagebreak

\end{widetext}
\clearpage

\bibliography{dSph_lines}

\end{document}